\documentclass[12pt]{article}
\include{epsf}
\usepackage{latexsym}
\usepackage{graphicx}

\begin{document}

\def\meanDelta33{\bar{\Delta}_{\bar{3}\bar{3}} }

\begin{titlepage}

%hep-ph/0112162
%\end{flushright}
\vskip1.8cm
\begin{center}
{\LARGE
Color Superconductivity \\
with Non-Degenerate Quarks      
\vskip0.6cm
}         
\vskip1.5cm
{\Large Todd Fugleberg} 
\vskip0.5cm
        Nuclear Theory Group \\
        Brookhaven National Laboratory \\
 Upton, NY, 11973-5000 \\ 
        {\small e-mail:
fugle@bnl.gov}\\
\vskip1.0cm
%PACS numbers: 
\begin{center}
BNL-NT-01/31
\end{center}

{\Large Abstract\\}
\end{center}
\parbox[t]{\textwidth}
{In dense quark matter, the response of the color superconducting gaps
to a small variation, $\delta\mu$, in the chemical potential of the
strange quark was studied.  The approximation of three massless
flavors of quarks and a general ansatz for the color flavor structure
of the gap matrix was used.  The general pole structure of the
quasi-particle propagator in this ansatz is presented.  The gap
equation was solved using both an NJL interaction model and
perturbative single gluon exchange at moderate densities and results
are presented for varying values of $\delta\mu$.  Quantitative and
qualitative differences in the dependence of the gaps on $\delta\mu$
were found.}

\vspace{1.0cm}

\end{titlepage}

\section{Introduction}

The physics of strongly interacting matter at high densities and low
temperatures has been the subject of much research in recent years.
It has long been known\cite{Barrois_BailinLove} that at sufficiently
high densities a system of quarks should form a condensate of Cooper
pairs which breaks the $SU_C(3)$ symmetry and becomes a color
superconductor.  The formation of the condensates leads to gaps in the
quasi-particle mass spectrum.  The authors of
\cite{Barrois_BailinLove} estimated that the gaps were of the order of
$\Delta\sim 10^{-3}\mu$ where $\mu$ is the quark chemical
potential. Recently it was shown at realistic values of $\mu$ in an
instanton induced NJL model that gaps of the order of $\mu$ could be
obtained\cite{ARW_RSSV}.  This stimulated a great deal of research\footnote{For
extensive references see the review articles \cite{Reviews}.} in
the ensuing years and has resulted in a proliferation of predicted
superconducting ground states.  These states may be realized in the
cores of neutron stars and lead to observable
effects\cite{CCSCCompact,Carter_Blinking}.

It is widely accepted \cite{CFL,CFL_accepted} that the color superconducting ground state at
asymptotic densities is the Color Flavor Locked (CFL) state\cite{CFL}:
\begin{equation}
\langle q_\alpha^i C \gamma^5 q_\beta^j  \rangle=  \Delta_{\bar{3}\bar{3}} 
(\delta^i_\alpha \delta^j_\beta-\delta^i_\beta\, \delta^j_\alpha)
+\Delta_{66}
(\delta^i_\alpha \delta^j_\beta+\delta^i_\beta\, \delta^j_\alpha),
\end{equation}
where the Greek indices are color indices, the Latin indices are
flavor indices and the $\bar{3}$ and $6$ subscripts refer to anti-triplet
or sextet configurations in color and flavor spaces respectively.
At lower densities ($\mu\sim m_s^2/4 \Delta$), it is likely that the ground
state is a superconducting state involving the condensation
of Cooper pairs in the $u-d$ sector only (2SC).  Finally at still lower
densities the favored ground state will be ordinary hadronic matter.
%, possibly accompanied by an independent $\bar{s}s$ condensate.
Two new phases have recently been predicted: Crystalline Color Superconductivity
\cite{Crystalline} and CFL with meson condensation\cite{Stress}.
These predictions, while not necessarily at odds with one another,
indicate that the transition region between the CFL state and hadronic
matter is not completely understood.

Including the strange quark mass in the gap equation with no 
approximations introduces two sets of complications: 1) massive quarks
means that there are 4 new types of Dirac structures allowed for the
condensates, and coupling between the corresponding gaps; 2) the fact that the
strange quark is different from the other quarks means that
gaps involving the strange quark should be different from those
with zero strangeness.

In order to understand the implications of these two complications it
is useful to separate them and understand them individually
before tackling the full problem.  Therefore, in this paper, all three
quarks are treated as massless and the strange quark is distinguished
by giving it a different chemical potential ($\mu_s=\mu-\delta\mu$)
than the $u$ and $d$ quarks.  This is motivated by observing\cite{Unlocking_ABR} that
for fixed Fermi energy, the Fermi momentum of a massive particle is
less than the Fermi momentum of a massless particle:
\begin{eqnarray}
E_F=p_{F0}=\mu_0 ~~~~~~~~~~~~~~~~~~~~~~~~~~~~~~~~&& \!\!\!\!\!\!\mbox{for massless particles,} \\
E_{F}=\sqrt{p_{Fm}^2+m^2}  \approx p_{Fm} + \frac{1}{2}\frac{m^2}{p_{Fm}}=\mu_0
&& \!\!\!\!\!\!\mbox{for massive particles.}
\end{eqnarray}
%This approach is motivated by the
%way $\mu$ enters the bare particle propagator in the combination:
%\begin{equation}
%(k_0 - \mu)=(\sqrt{k^2+m^2}\pm \mu)\approx \left(k + \frac{1}{2}\frac{m^2}{k} \pm \mu\right)
%\end{equation}
This approximately corresponds to a shift in the Fermi momentum
or the chemical potential:
\begin{equation}
\mu_m=p_{Fm}\approx \mu_0 -\frac{1}{2}\frac{m^2}{p_{Fm}} \equiv \mu_0 -\delta\mu.
\end{equation}
This approach is similar to that taken in \cite{SW_Description} where
this approach was studied using a four fermion interaction (NJL) model.  
Qualitatively similar results to those of
\cite{SW_Description} were obtained in \cite{Unlocking_ABR} using an
NJL interaction and a different prescription for inclusion
of strange quark mass effects.  In this paper an NJL model
is considered in Section 3 for comparison to previous results and to
illustrate the contrast with the results of the subsequent section.  In
section 4, perturbative single gluon exchange is considered for the interaction of
the quarks.  This approach is valid in the high density regime and
gives an approximate model in the moderate density regime which is
distinct from and can be compared to the NJL interaction model.

Understanding the effects of a shift in the chemical potential of one
of three quark flavors is another motivation for this research.  The
papers \cite{Crystalline,Asym_Bedaque,Enforced,Diagrammatic} also deal
with shifts in the chemical potential in an NJL model, mostly in the
two flavor case and find some interesting effects.  The three flavor
case was discussed in \cite{Crystalline} and \cite{Enforced} but was
not analyzed in detail.  In \cite{Enforced} it was predicted
that the number densities of the 3 quark flavors are constrained
to be equal up to $\delta\mu=\Delta_{\bar{3}\bar{3}}/\sqrt{2}$, at
which point there is a first order transition to some less symmetric
phase of quark matter.  It would be interesting to
test this prediction using perturbation theory.  Finally in \cite{Opening}
first order perturbation theory is used to study the non-degenerate two flavor case.  In
this work, first order perturbation theory is applied in the non-degenerate three flavor case.

Perturbation theory applied to Color Superconductivity has been the
subject of much important research\footnote{See for example
\cite{1GExchange,RD_Weak,Perturbation} and references in \cite{Reviews}.}.  It
has not previously been applied to the case of color superconductivity
with three flavors of non-degenerate quarks.

Recently numerical results were obtained\cite{Kazu} in the 2 flavor
case in perturbation theory with single gluon exchange and 
massless and degenerate quarks.  The $\delta\mu=0$ results presented in this
paper for perturbative single gluon exchange with three flavors are
qualitatively similar and can be considered as independent
confirmation of \cite{Kazu} as well as an extension of them to three flavors.

The fact that the three flavor case is different from the two flavor
case for non-degenerate flavors should be
emphasized at this point.  This is clear from the NJL
analyses of the two different cases.  In
\cite{Crystalline,Diagrammatic} the gap is shown to be independent
of $\delta\mu$ up to some critical value.  In
\cite{Unlocking_ABR,SW_Description}, $\delta\mu$ has a significant
effect on the pattern of condensation even at small values of
$\delta\mu$.  This difference is simply a result of the fact that the
algebra of the generators of $SU(2)$ is much simpler than the algebra
of the generators of $SU(3)$.  The important point that this illustrates
is that results from the two flavor case may not tell the full story in
the three flavor case.

The goals of this work are threefold. The first goal is to understand the 
effects of the non-degeneracy of the strange quark in a simplified model of QCD
as a first step toward understanding the effect of the strange quark 
mass in a more complete way.  The second goal of this work is to understand 
the effects of a small shift in the chemical potential on solutions
of the gap equation for comparison with results in 
other approaches to the problem.  The third goal is to determine
how first order perturbative results compare to results obtained using the NJL 
interaction model.

Results presented in this paper are the poles of the quasiparticle
propagator in this approach in a more general ansatz than in
\cite{SW_Description} and in a different model than \cite{{Unlocking_ABR}}.  
Numerical solutions for the gaps at moderate densities using 
both an NJL interaction model and perturbative single gluon exchange are presented to 
illustrate the effect of $\delta\mu$ and the qualitative and quantitative 
differences in the results.

\section{Gap Equation}

The gaps resulting from the formation of a color superconducting
ground state
can be determined by solving the mean field gap equation\cite{RD_superfluid}:
\begin{equation}
\Delta(k)= - i g^2 \int \frac{d^4q}{(2 \pi)^4} \sum_{A,B} D^{\mu\nu}_{AB}(k-q)
\gamma_\mu (\frac{\lambda^A_c}{2})^T G^-_0(q) \Delta(q) G^+(q) \gamma_\nu 
\frac{\lambda^B_c}{2}
\label{GapEquation1}
\end{equation}
which is specialized to an interaction with the color structure
of single gluon exchange.  $D^{\mu\nu}_{AB}(k-q)$ is the gluon propagator,
$\lambda^A_c$ are the Gell-Mann matrices, $G^-_0(q)$ is the bare antiquark propagator,
$G^+(q)$ is the full quasiparticle propagator and $\Delta(k)$ is the gap matrix.

%Notice that the object between the interaction vertices with the 
%gluon is the (2,1) component of full Nambu-Gorkov propagator,${\cal S}$.

The gap matrix is a matrix in color, flavor and spinor space that 
contains all the specific gaps. Restricting consideration
to spin zero gaps in the massless case the 
Dirac structure of the gap matrix can be written in a basis of 
four projectors\cite{RD_superfluid}:
\begin{equation}
P^e_h(\vec{k})= \Lambda^e(\vec{k})P_h(\hat{k})=
\left(\frac{1+e\,\gamma_0\vec{\gamma} \cdot {\vec k}}{2} \right)
\left(\frac{1+e\,\gamma_5\gamma_0\vec{\gamma} \cdot {\vec k}}{2} \right), ~~~~ e,h=\pm 1
\end{equation}
which are products of projectors onto positive and negative energy states and
positive and negative helicity states.

The color flavor structure of all objects will be given as $9\times 9$ matrices
consisting of a $3\times 3$ matrix in flavor space whose components
are $3\times 3$ matrices in color space:
%\begin{equation}
%M^{ij}_{\alpha\beta}=
%\left(
%\begin{array}{lcr}
%M^{11}_{\alpha\beta} & M^{12}_{\alpha\beta}& M^{13}_{\alpha\beta} \\
%M^{21}_{\alpha\beta} & M^{22}_{\alpha\beta}& M^{23}_{\alpha\beta} \\
%M^{31}_{\alpha\beta} & M^{32}_{\alpha\beta}& M^{33}_{\alpha\beta}
%\end{array} \right).
%\end{equation}
\begin{eqnarray}
\!\!\!\!\!\!\!\!\!\!\!\!\!\!\!\!\!\!j=u~~~~ & ~~j=d~~ & ~~~j=s~~~~~
\end{eqnarray}

\vspace{-0.4in}
\begin{eqnarray}
\hspace{-1in}
\begin{array}{c}
~\\
i=u \\
~\\
~\\
i=d \\
~\\
~\\
i=s \\
~\\
\end{array}
\left(
\begin{array}{ccccccccc}
\Box & \Box & \Box & \Box & \Box & \Box & \Box & \Box & \Box \\
\Box & \Box & \Box & \Box & \Box & \Box & \Box & \Box & \Box \\
\Box & \Box & \Box & \Box & \Box & \Box & \Box & \Box & \Box \\
\Box & \Box & \Box & \alpha, & \beta & = & \Box & \Box & \Box \\
\Box & \Box & \Box & (r, & g, & b) & \Box & \Box & \Box \\
\Box & \Box & \Box & ~~ & ~~ & ~~ & \Box & \Box & \Box \\
\Box & \Box & \Box & \Box & \Box & \Box & \Box & \Box & \Box \\
\Box & \Box & \Box & \Box & \Box & \Box & \Box & \Box & \Box\\
\Box & \Box & \Box & \Box & \Box & \Box & \Box & \Box & \Box
\end{array}
\right)
\nonumber
\end{eqnarray}

The most general gap matrix can be written as:
\begin{equation}
\Delta(\vec{k})=\sum_{e,h=\pm 1} \left(\Delta^{\alpha\beta}_{ij}\right)^e_h
 P^e_h(\vec{k}).
\end{equation}

The gluon propagator, $D^{\mu\nu}_{AB}$, is the object that
differs between the NJL model and perturbation theory.
In the NJL model the interaction is modeled by a 4 fermion interaction
and the gluon propagator is replaced by an identity matrix.
This constitutes a low energy effective model of the interaction.
In perturbative single gluon exchange $D^{\mu\nu}_{AB}$
is the bare gluon propagator.  Perturbation theory is definitely valid at high energies
and densities.  At moderate densities perturbation theory can be considered a model
for the quark interactions.

The bare propagator for the a quark of flavor $f$ is given by:
\begin{equation}
\left(G^\pm_0(k)\right)_f
=\left(\gamma^\nu k_\nu \pm \mu_f \gamma_0 \right)^{-1}
=\left((k_0\pm\mu_f)\gamma_0-\vec{\gamma}\cdot \vec{k} \right)^{-1}
\end{equation}
If all $\mu_f$ are equal the particles are degenerate and the
bare propagator is the identity on the color-flavor space.  In the
case considered here the bare antiquark propagator can be written
in the form:
\begin{eqnarray}
G^-_0&=&\left((k_0-\mu-\delta\mu\, P^{(s)}_f) \gamma^0-|k|\,\vec{\gamma} \cdot
\hat{k} \right)^{-1} 
%= \gamma^0\;\left((k_0-\mu-\delta\mu\, P^{(s)}_f) +k\,\gamma^0\gamma \cdot \hat{k} \right)^{-1} 
\label{BarePropagator}\\ &=&
\gamma^0\;\sum_{e,h} \frac{\left(P^{(u)}_f+P^{(d)}_f\right)
P^e_h(k)}{(k_0-\mu ) + e \, |k|} + \gamma^0\;\sum_{e,h} \frac{P^{(s)}_f
P^e_h(k)}{(k_0-\mu - \delta\mu \,
\delta_{i3}) + e \, |k|} \nonumber 
\end{eqnarray}
where the $P^{(i)}$ are projectors onto a specific flavor sub-space, ie:
%\begin{equation}
%P^{(u)}_f=
%\left(
%\begin{array}{lcccccccr}
%1 & 0 & 0 & 0 & 0 & 0 & 0 & 0 & 0\\
%0 & 1 & 0 & 0 & 0 & 0 & 0 & 0 & 0\\
%0 & 0 & 1 & 0 & 0 & 0 & 0 & 0 & 0\\
%0 & 0 & 0 & 0 & 0 & 0 & 0 & 0 & 0\\
%0 & 0 & 0 & 0 & 0 & 0 & 0 & 0 & 0\\
%0 & 0 & 0 & 0 & 0 & 0 & 0 & 0 & 0\\
%0 & 0 & 0 & 0 & 0 & 0 & 0 & 0 & 0\\
%0 & 0 & 0 & 0 & 0 & 0 & 0 & 0 & 0\\
%0 & 0 & 0 & 0 & 0 & 0 & 0 & 0 & 0 
%\end{array} \right)
%\end{equation}
\begin{equation}
P^{(u)}_f=
\left(
\begin{array}{lcr}
{\bf 1} & {\bf 0} & {\bf 0} \\
{\bf 0} & {\bf 0} & {\bf 0} \\
{\bf 0} & {\bf 0} & {\bf 0} 
\end{array} \right)
\end{equation}
where bold font indicates $3\times3$ matrices in color space,
and similarly for $P^{(d)}_f$ and $P^{(s)}_f$.

The quasiparticle (quasi-antiparticle) propagator is determined
from the bare particle and antiparticle propagators and the gap matrix
by the relation\cite{RD_superfluid}:
\begin{equation}
G^\pm\equiv\left\{ [G^\pm_0]^{-1}-\Delta^\mp G^\mp_0 \Delta^\pm\right\}^{-1}.
\label{quasiparticle_propagator}
\end{equation}
where $\Delta^+\equiv \Delta$ and $\Delta^-\equiv \gamma_0 \Delta^+ \gamma_0$.

The ansatz for the color-flavor structure of the
gap matrix used in this research is:
\begin{eqnarray}
\hspace{-0.2in}\Delta^{\alpha\beta}_{ij}=\left(   
\begin{array}{ccccccccc}
\frac{a+h+e}{2} & ~0~ & ~0~ & ~0~ & \frac{a+h-e}{2}   & ~0~ & ~0~ & ~0~ & ~c~ \\
0   & 0 & 0 & e & 0   & 0 & 0 & 0 & 0 \\
0   & 0 & 0 & 0 & 0   & 0 & f & 0 & 0 \\
0   & e & 0 & 0 & 0   & 0 & 0 & 0 & 0 \\
\frac{a+h-e}{2}   & 0 & 0 & 0 & \frac{a+h+e}{2} & 0 & 0 & 0 & c \\
0   & 0 & 0 & 0 & 0   & 0 & 0 & f & 0 \\
0   & 0 & f & 0 & 0   & 0 & 0 & 0 & 0 \\
0   & 0 & 0 & 0 & 0   & f & 0 & 0 & 0 \\
c   & 0 & 0 & 0 & c   & 0 & 0 & 0 & a-h 
\end{array}
\right)  
\label{Color_Flavor_Ansatz}
\end{eqnarray}
where:
\begin{eqnarray}
\frac{a+h-e}{2}&=&\Delta_{66}(ud) +\Delta_{\bar{3}\bar{3}}(ud), \\
e&=&\Delta_{66}(ud)-\Delta_{\bar{3}\bar{3}}(ud), \\
c&=&\Delta_{66}(us)+\Delta_{\bar{3}\bar{3}}(us), \\
f&=&\Delta_{66}(us)-\Delta_{\bar{3}\bar{3}}(us),\\
a-h&=& \Delta_{66}(ss)
\end{eqnarray}
The form of this ansatz is essentially the same as in \cite{Unlocking_ABR}
with minor differences to simplify the poles of the quasi-particle
propagator.  The correspondence between the different conventions is:
\begin{eqnarray}
b=\frac{a+h-e}{2} ~&~ d=a-h 
\end{eqnarray}
and $e$,$c$,$f$ unchanged.

In the case of degenerate color-flavor locking:
\begin{eqnarray}
c=\frac{a+h-e}{2} ~~&~~ e=f ~~ &~~ a-h=\frac{a+h+e}{2}
\end{eqnarray}

In order to simplify the calculation, following the example of 
\cite{Unlocking_ABR}, the ansatz (\ref{Color_Flavor_Ansatz}) is rotated 
to a different basis.  The columns (rows) are referred to by the
combination of color and flavor indices, ($i,\alpha$).\footnote{Therefore for 
instance the 2nd column is the $(i,\alpha)=(u,g)$ column.}
The $(u,r)$ and $(d,g)$ basis vectors are rotated to
new basis vectors $1/\sqrt{2} \left( (u,r) \mp (d,g) \right)$ and all
the other basis vectors are left unchanged\footnote{This means for example that
the new first column is $1/\sqrt{2} \left( (u,r) - (d,g) \right)$ and
similarly for the fifth column.}.  
%This change of basis simplifies the results
%a little and does not complicate the calculation that much.  
Note that this change of basis leaves $\left(P^{(u)}_f+P^{(d)}_f\right)$ and
$P^{(s)}_f$ operators unchanged which means that the bare propagator
has the same form in this new basis.
In this basis the gap matrix now has the form:
\begin{equation}
\Delta^{\alpha\beta}_{ij}=\left(   
\begin{array}{ccccccccc}
e & 0 & 0 & 0 & 0   & 0 & 0 & 0 & 0 \\
0   & 0 & 0 & e & 0   & 0 & 0 & 0 & 0 \\
0   & 0 & 0 & 0 & 0   & 0 & f & 0 & 0 \\
0   & e & 0 & 0 & 0   & 0 & 0 & 0 & 0 \\
0   & 0 & 0 & 0 & a+h & 0 & 0 & 0 & \sqrt{2}c \\
0   & 0 & 0 & 0 & 0   & 0 & 0 & f & 0 \\
0   & 0 & f & 0 & 0   & 0 & 0 & 0 & 0 \\
0   & 0 & 0 & 0 & 0   & f & 0 & 0 & 0 \\
0   & 0 & 0 & 0 & \sqrt{2}c   & 0 & 0 & 0 & a-h
\end{array} 
\right) 
\label{ansatz} \nonumber
\end{equation}
Of course all other objects with color or flavor structure must be transformed to the same basis.
These details are only included in order that the quasiparticle propagator can be presented
below in a fairly concise way.

Up until this point no assumptions about the Dirac
structure of the gap matrix has been made.  For ease of calculation
in the perturbative analysis
%for purposes of clarity I will make some assumptions
%about the Dirac structure of the condensates although this is not
%required and the calculation could be carried out in full generality.
the gap matrix is assumed to have the usual form:
\begin{equation}
\Delta(k)=\langle q_\alpha^i(-k) C  \gamma_5 q_j^\beta(k) \rangle= 
\Delta^{\alpha\beta}_{ij} 
\sum_{e,h=\pm 1} e\, h \; P^e_h(k)
\label{Dirac_ansatz}
\end{equation}

Substituting (\ref{Dirac_ansatz}) and (\ref{ansatz}) into
 (\ref{quasiparticle_propagator}), the quasiparticle
propagator is:
\begin{equation}
G^+=
\left(
\begin{array}{ccccccccc}
E   & 0 & 0 & 0 & 0   & 0 & 0 & 0 & 0 \\
0   & E & 0 & 0 & 0   & 0 & 0 & 0 & 0 \\
0   & 0 & F_1 & 0 & 0   & 0 & 0 & 0 & 0 \\
0   & 0 & 0 & E & 0   & 0 & 0 & 0 & 0 \\
0   & 0 & 0 & 0 & A   & 0 & 0 & 0 & C \\
0   & 0 & 0 & 0 & 0   & F_1 & 0 & 0 & 0 \\
0   & 0 & 0 & 0 & 0   & 0 & F_2 & 0 & 0 \\
0   & 0 & 0 & 0 & 0   & 0 & 0 & F_2 & 0 \\
0   & 0 & 0 & 0 & C   & 0 & 0 & 0 & B
\end{array}
\right),
\label{PropagatorMatrix}
\end{equation}
where:
\begin{eqnarray}
E&=& \left[\frac{q_0+l}{\left(q_0-l\right)
\left(q_0+l\right)-e^2}\right] \; \Lambda^+ + \left[~ l 
\rightarrow -l' ~\right] \Lambda^-  \\
%-\frac{1}{\left(q_0-(|q|+\mu)\right))
%\left(q_0+(|q|+\mu)\right))-e^2} \; \nonumber\\
F_1&=& \left[\frac{q_0+l-\delta\mu}{\left(q_0-l\right)
\left(q_0+l-\delta\mu\right)-f^2} \right] \; \Lambda^+  + \left[~ l 
\rightarrow -l' ~\right] \Lambda^-  \\
F_2&=& \left[\frac{q_0+l}{\left(q_0-l+\delta\mu\right)
\left(q_0+l\right)-f^2} \right] \; \Lambda^+ + \left[~ l 
\rightarrow -l' ~\right] \Lambda^-  \\
%&-& \frac{q_0-(|q|+\mu)-\delta\mu}{\left(q_0-(|q|+\mu)\right)-\delta\mu)
%\left(q_0+(|q|+\mu)\right))-f^2} \;F_1\\
%&-& \frac{q_0-(|q|+\mu)}{\left(q_0-(|q|+\mu)\right))
%\left(q_0+(|q|+\mu)\right)+\delta\mu)-f^2} \;F_2 \nonumber\\
\!A&=&\!\!-\left[\frac{(a-h)^2(l+q0)+(2 c^2+(l+q_0)(l-q_0-\delta\mu))(l+q_0-\delta\mu)))}
{\left(q_0^2-\epsilon_+(a,h,c,l,\delta\mu)^2\right)
\left(q_0^2-\epsilon_-(a,h,c,l,\delta\mu)^2\right)}\right]  \Lambda^+
 \nonumber\\
&-& \left[~ l 
\rightarrow -l' ~\right] \Lambda^-  \\
\!B&=&\!\!-\left[\frac{(a+h)^2(l+q0-\delta\mu)+(2 c^2+(l-q_0)(l+q_0-\delta\mu))(l+q_0)))}
{\left(q_0^2-\epsilon_+(a,h,c,l,\mu,\delta\mu)^2\right)
\left(q_0^2-\epsilon_-(a,h,c,l,\mu,\delta\mu)^2\right)}\right]
\Lambda^+\nonumber\\
&-& \left[~ l 
\rightarrow -l' ~ \right] \Lambda^-  \\
C&=&\left[\frac{\sqrt{2} c \left( a (2l + 2q_0-\delta\mu)- h \delta\mu  \right)}
{\left(q_0^2-\epsilon_+(a,h,c,l,\mu,\delta\mu)^2\right)
\left(q_0^2-\epsilon_-(a,h,c,l,\mu,\delta\mu)^2\right)} \right]\Lambda^+ \nonumber\\
&+& \left[~ l 
\rightarrow -l' ~ \right] \Lambda^-
\end{eqnarray}
and:
\begin{eqnarray}
&\!&\!\!\!\!\epsilon_\pm(a,h,c,l,\delta\mu)^2=l(l-\delta\mu)
+\delta\mu^2+a^2+2 c^2+h^2 \\
&~&~\pm\sqrt{4  \, a^2 (2 c^2+h^2)+2  \,a \, h (2 l -\delta\mu) 
\delta\mu + \frac{1}{4}
 \delta\mu^2 (8 c^2+ (2 l -\delta\mu)^2)}\;\;,  \nonumber
\end{eqnarray}
using the definitions:
%\begin{equation}
%l=(|q|-\mu),
%\end{equation}
\begin{eqnarray}
l&=&(|q|-\mu), \\
l'&=&(|q|+\mu), 
\end{eqnarray}
which are the only ways that the functions depends on $|q|$ and $\mu$.

The form (\ref{PropagatorMatrix}) of the propagator is consistent
with expectations.  The strange quark has been
distinguished from the up and down quarks.  The ansatz still
has flavor locked to color and therefore the blue quark has
been distinguished from the red and green quarks. 
The $(u,g)$ and $(d,r)$ quarks propagators are unchanged.
The $(u,b)$ and $(d,b)$ propagators are affected in exactly
the same way.  The $(s,r)$ and $(s,g)$ propagators
are also affected in exactly the same way.  If one rotated back to the
original basis one would find the $(u,r)$ and $(d,g)$ propagators are
also affected in exactly the same way.
%This structure for the propagator makes sense because the $(u,g)$ and $(d,g)$
%(2nd and 4th) columns which involve only the first two colors and flavors are identical.
%The $(u,b)$ and $(d,b)$ columns (3rd and 6th) columns which involve the one of the 
%first two identical flavors and the third color are identical.  The $(s,r)$ and $(s,g)$
%(7th and 8th) which involves the third flavor and the first two colors are identical.
The propagator could be completely diagonalized to separate the poles 
which would simplify the calculation in some ways but would complicate it in others,
so this form is used in this work.

The poles of the quasi-particle propagator are therefore:
\begin{eqnarray}
q_0=\pm \sqrt{l^2+e^2}~~~~~~~~~~~~~~~~~~~~~~~~~~~~~~~~~\mbox{degeneracy}=3, 
\label{pole1}
\end{eqnarray}
\begin{eqnarray}q_0=\frac{\delta\mu}{2} \pm \sqrt{(l-\frac{\delta\mu}{2})^2+f^2}
~~~~~~~~~~~~~~~~~~\mbox{degeneracy}=2 ,
\label{pole2}
\end{eqnarray}
\begin{eqnarray}
q_0=-\frac{\delta\mu}{2} \pm \sqrt{(l-\frac{\delta\mu}{2})^2+f^2}
~~~~~~~~~~~~~~~~\mbox{degeneracy}=2 , 
\label{pole3}
\end{eqnarray}
\begin{eqnarray}
q0=\pm \epsilon_-(a,h,c,l,\delta\mu)~~~~~~~~~~~~~~~~~~~~~~~~~\mbox{degeneracy}=1 ,
\label{pole4}
\end{eqnarray}
\begin{eqnarray}
q0=\pm \epsilon_+(a,h,c,l,\delta\mu)~~~~~~~~~~~~~~~~~~~~~~~~~\mbox{degeneracy}=1.
\label{pole5}
\end{eqnarray}
%\begin{eqnarray}
%q_0&=&\pm\left[l(l-\delta\mu) +\delta\mu^2+a^2+ a_{12}^2 2 c^2+h^2
%\right. \label{pole4} \\
%\pm &\!& \!\!\!\!\!\!\!\!\!\! \left. 
%\sqrt{4  \, a^2 (2 c^2+h^2)+2  \,a \, h (2 l -\delta\mu) 
%\delta\mu + \frac{1}{4}
% \delta\mu^2 (8 c^2+ (2 l -\delta\mu)^2)} \; \right]^{1/2}  ,
% \nonumber
%\end{eqnarray}
It can be shown that in the case where the sextet
gaps\footnote{The sextet gaps are much
smaller than the anti-triplet gaps.} are neglected, the poles of this
propagator reduce to those given in \cite{SW_Description}.  Further
the first eight of the poles reduce to the octet poles and the last
pole reduces to the singlet pole at $\delta\mu=0$.  The poles obtained
in this model are based on a different model than
\cite{Unlocking_ABR}.
% and are not obviously the same although one
% would expect them to lead to the same physics.

Substituting the ansatz (\ref{ansatz}), the solution for $G^+$(\ref{PropagatorMatrix}),
and the bare propagator into the gap equation (\ref{GapEquation1}) 
one obtains a matrix gap equation.

\begin{eqnarray}
\Delta(k)&=&-i g^2 \int \frac{d^4q}{(2 \pi)^4}  D^{\mu\nu}(k-q)
\gamma_\mu \gamma^0 \gamma^5\gamma^0\gamma\cdot\hat{k}\\
&&\left(\Lambda^+(q) M(a,h,c,l,q_0,\delta\mu)-\Lambda^-(q)
M(a,h,c,-l',q_0,
\delta\mu)\right)\gamma^0
\gamma_\nu \nonumber \\
&=&-i g^2 \int \frac{d^4q}{(2 \pi)^4}  D^{\mu\nu}(k-q)  \gamma^5
\gamma_\mu \gamma\cdot\hat{k}\gamma^0\\
&&\left(\Lambda^-(q) M(a,h,c,l,q_0,\delta\mu)-\Lambda^+(q)
M(a,b,a_{12},-l',q_0,
\delta\mu)\right) \gamma_\nu \nonumber
\label{MatrixGapEquation}
\end{eqnarray}
where 
%we must use $\lambda^A$ rotated to the basis we are working in and 
$D^{\mu\nu}_{AB}(k-q)=\delta_{AB} D^{\mu\nu}(k-q)$ has been used,
%which is consistent with single gluon exchange.  Finally we have 
the index $A$ has been summed over and the $\lambda$ matrices have
been multiplied through.

The matrix $M(a,h,c,l,q_0,\delta\mu)$
has the form:
\begin{equation}
M=\left(   
\begin{array}{ccccccccc}
M_{11} & 0 & 0 & 0 & 0   & 0 & 0 & 0 & 0 \\
0   & 0 & 0 & M_{11} & 0   & 0 & 0 & 0 & 0 \\
0   & 0 & 0 & 0 & 0   & 0 & M_{37} & 0 & 0 \\
0   & M_{11} & 0 & 0 & 0   & 0 & 0 & 0 & 0 \\
0   & 0 & 0 & 0 & M_{55} & 0 & 0 & 0 & M_{59} \\
0   & 0 & 0 & 0 & 0   & 0 & 0 & M_{37} & 0 \\
0   & 0 & M_{73} & 0 & 0   & 0 & 0 & 0 & 0 \\
0   & 0 & 0 & 0 & 0   & M_{73} & 0 & 0 & 0 \\
0   & 0 & 0 & 0 & M_{95}   & 0 & 0 & 0 & M_{99}
\end{array}
\right)
\end{equation}
which should be compared to the form of the ansatz (\ref{ansatz}).  In
general $M_{73}\neq M_{37}$ and $M_{95}\neq M_{59}$ which can be
shown to be inevitable if one examines the product $G^-_0 \Delta^+
G^+$ carefully.  However, the terms which spoil the symmetry of these
components can be shown to vanish under the $q_0$ integration over the
interval $(-\infty,\infty)$.  This is immediately clear in the NJL
model and follows with a little more care in perturbation theory.  
Therefore the matrix gap equation does close under this ansatz.

Everything discussed so far follows without specification of the
interaction model.  In the following section the 
gap equations in the NJL interaction model will be given for 
the ansatze above.  Solutions for the gaps will be presented in order to compare to
previous results and to highlight the different behavior of the
gaps using perturbation theory.

\section{NJL Interaction Model}

In an NJL model the fermion interaction is taken to be:
\begin{equation}
g^2 D^{\mu\nu}_{AB} \rightarrow  G g^{\mu\nu} \delta_{AB}
\end{equation}
which is the same form as the interaction model of
\cite{Unlocking_ABR} and the gap equation becomes:
\begin{eqnarray}
\Delta(k)&=&  -i G \int \frac{d^4q}{(2 \pi)^4} 
 \gamma^5\gamma^\nu \gamma\cdot\hat{k}\gamma^0\\
&&
\left(\Lambda^-(q) M(a,h,c,l,q_0,\delta\mu)-\Lambda^+(q)
M(a,h,c,-l',q_0.
\delta\mu)\right) \gamma_\nu \nonumber
\end{eqnarray}

Using the relation: 
$\gamma^\nu \gamma^0 \gamma^ i \gamma_\nu=0$, multiplying by $\gamma^5$ and tracing over
Dirac indices one obtains:
%\begin{equation}
%\Delta= 4 G \int \frac{d^4q}{(2 \pi)^4}  \gamma^5
%\left( M(a,h,c,l,q_0,\delta\mu)+
%M(a,h,c,-l',q_0,
%\delta\mu)\right)
%\end{equation}
%
%This is still a matrix in Dirac space but at t multiply
%by $\gamma^5$ on both sides and trace over Dirac indices to obtain:
\begin{equation}
\Delta^{\alpha\beta}_{ij}=  4 i G \int \frac{d^4q}{(2 \pi)^4}  
\left( M(a,h,c,l,q_0,\delta\mu)+M(a,h,c,-l',q_0,\delta\mu)\right)
\end{equation}
where the right hand side of the equation does not depend on $k$ anymore
and the functional dependence on the left hand side has been dropped.
Equating coefficients of matrices on each side produces a set of 
coupled gap equations:
\begin{eqnarray}
&\!\!\!\!\!\!\!\!\!\!&e=4 i G\int \frac{d^4q}{(2 \pi)^4}\left[
\frac{5 e}{12(e^2+l^2-q_0^2)} \right. \label{NJL_gap_equation_e} \\ 
&\!\!\!\!\!\!\!\!\!\!&
+\left.
\frac{2(a-h)c^2-(a+h)\left((a-h)^2 +(l-\delta\mu)^2-q_0^2 \right)}
{4\left(q_0^2-\epsilon_+^2\right)
\left(q_0^2-\epsilon_-^2\right)}
   \right]\nonumber
+ \left[ l \rightarrow -l' \right]\\ 
&\!\!\!\!\!\!\!\!\!\!&
f=4 i G \int \frac{d^4q}{(2 \pi)^4} \left[
\frac{c(a^2 - 2c^2 - h^2 - l^2 + q0^2 + l \delta\mu)}
{2\left(q_0^2-\epsilon_+^2\right)
\left(q_0^2-\epsilon_-^2\right) }
 \right.\label{NJL_gap_equation_f}\\ 
&\!\!\!\!\!\!\!\!\!\!&
+\left.
\frac{f}{12(f^2 + (l + q0) (l - q0 - \delta\mu))} +
\frac{f}{12(f^2 + (l - q0) (l + q0 - \delta\mu))}
 \right]+ \left[ l \rightarrow -l' \right]
\nonumber \\ 
&\!\!\!\!\!\!\!\!\!\!&
a=4 i G\int \frac{d^4q}{(2 \pi)^4}\left[-\frac{3 e}{8(e^2+l^2-q_0^2)}~~ +\right.\label{NJL_gap_equation_a}\\ &\!\!\!\!\!\!\!\!\!\!&
\left.\frac{2(5a+3h)c^2+(a+h)(2l-\delta\mu)\delta\mu-(5a-3h)(l^2-q_0^2)-(a-h)(a+h)(5a+3h)}
{24\left(q_0^2-\epsilon_+^2\right)
\left(q_0^2-\epsilon_-^2\right)}
   \right]\nonumber\\ &\!\!\!\!\!\!\!\!\!\!&
+ \left[ l \rightarrow -l' \right]
\nonumber\\ 
&\!\!\!\!\!\!\!\!\!\!&
\\ 
&\!\!\!\!\!\!\!\!\!\!&
h=-4 i G\int \frac{d^4q}{(2 \pi)^4}\left[\frac{3 e}{8(e^2+l^2-q_0^2)}~~+\right.\label{NJL_gap_equation_h}\\ &\!\!\!\!\!\!\!\!\!\!&
\left.\frac{2(3a+5h)c^2-(a+h)(2l-\delta\mu)\delta\mu-(3a-5h)(l^2-q_0^2)-(a-h)(a+h)(3a+5h)}
{24\left(q_0^2-\epsilon_+^2\right)
\left(q_0^2-\epsilon_-^2\right)}
   \right]\nonumber\\ &\!\!\!\!\!\!\!\!\!\!&
+ \left[ l \rightarrow -l' \right]
\nonumber\\ 
&\!\!\!\!\!\!\!\!\!\!&
c=-4 i G\int \frac{d^4q}{(2 \pi)^4} \left[
\frac{c(a^2 - 2c^2 - h^2 - l^2 + q0^2 + l \delta\mu)}
{6\left(q_0^2-\epsilon_+^2\right)
\left(q_0^2-\epsilon_-^2\right) }
\right.\label{NJL_gap_equation_c}\\ 
&\!\!\!\!\!\!\!\!\!\!&
\left.+
\frac{f}{4(f^2 + (l + q0) (l - q0 - \delta\mu))} +
\frac{f}{4(f^2 + (l - q0) (l + q0 - \delta\mu))}
 \right]+ \left[ l \rightarrow -l' \right]
\nonumber
\end{eqnarray}

These gap equations include the terms $l \rightarrow -l^\prime$ which
correspond to the contribution of the anti-particle gap and do not
lead to a gap at the Fermi surface.  The anti-particle gap is not
necessarily small and in fact is equal to the particle gap in the NJL
model.  These terms are suppressed, however, by at least $1/\mu^2$ and
are usually neglected as is done here.  This is crucial because the antiparticle
gap is gauge dependent (see for example \cite{SW_Description}).
%  Further it was shown in
%\cite{MassTerms} for a few (?) physical quantities which might be expected to 
%depend on the anti-particle gap parameter that this dependence cancels out
%exactly.  For further comments see the next section.

Evaluation of the integrals on the right hand side of the gap
equations is facilitated by the analytic continuation, $q_0 \rightarrow -i\,q_4$.
The $q_4$ integral is done by contour integration
closing the contour in the upper half plane and picking up the poles, (\ref{pole1})-(\ref{pole5}),
which lie in the upper half plane after analytic continuation.  The angular
integrals can be done trivially reducing the right hand side of the
gap equation to an integration over $q(\equiv |\vec q|)$ which can be
done numerically.  This approach has the advantage that the
quasi-particles are automatically on the mass shell.

The range of integration for $q$ is not infinite 
since the NJL model is a four-fermion interaction model and must
have an UV cutoff.  In this work the integrals are simply regularized 
by the factor:
\begin{equation}
{\cal R}(q)=\frac{\Lambda^4}{(q^2+\Lambda^2)^2}.
\end{equation}
with $\Lambda=1000$ MeV chosen so that the magnitude of the results
in the NJL model are comparable to the perturbative results.

\begin{figure}[t]
\epsfysize=3in
\epsfbox[ 10 49 395 292]{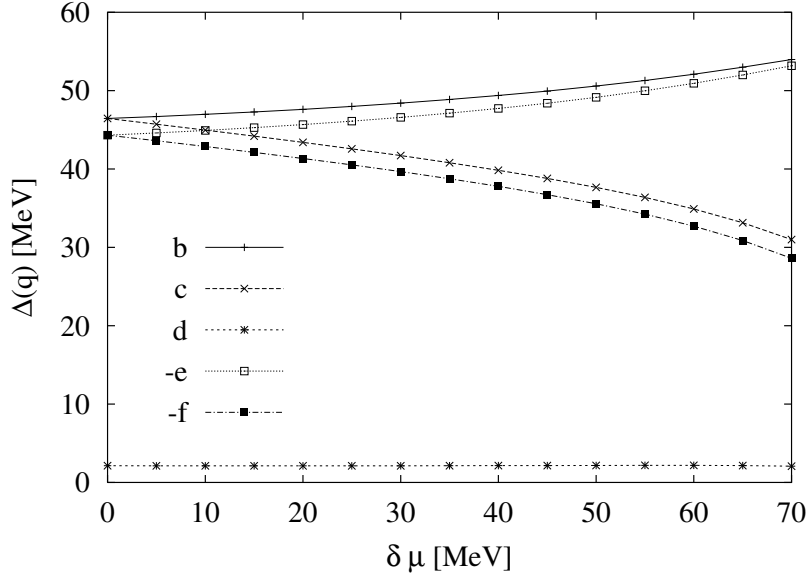}
   \caption{Numerical solutions for gaps as a function of $\delta\mu$ in the conventions
of \cite{Unlocking_ABR}.}
   \label{NJL_Figure_ABR}
\end{figure}

The coupled gap equations were solved iteratively for $\mu=500$ MeV using numerical
integration and the results are shown in Figure \ref{NJL_Figure_ABR},
after conversion to the convention of \cite{Unlocking_ABR} for
comparison. The effect of a shift in the strange quark
chemical potential is a disruption to the pairing of the heavy quark
with the light quarks in the dominant channel and a slightly smaller enhancement of
the pairing of the light quarks.  The results shown in
Fig. \ref{NJL_Figure_ABR} exhibit the same qualitative behavior as
shown in Figure 6 of \cite{Unlocking_ABR}.  Note that the range of
$\delta\mu$ shown in Fig. \ref{NJL_Figure_ABR} does not go all the way
up to the phase transition illustrated in \cite{Unlocking_ABR} and the
shifts seen in the results of this paper are larger than in \cite{Unlocking_ABR}.

\begin{figure}[ht]
\epsfysize=3in
\epsfbox[ 10 49 395 292]{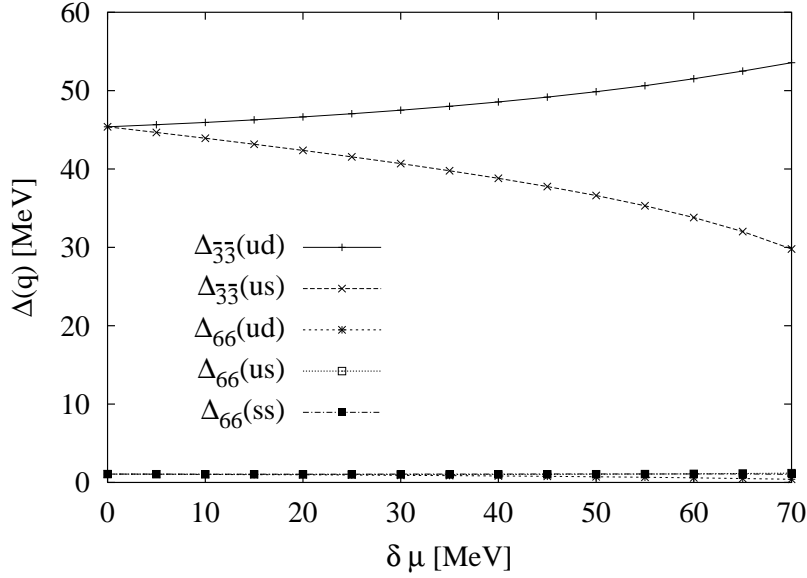}
   \caption{Numerical solutions for gaps as a function of $\delta\mu$ with specific color
flavor symmetries.}
   \label{NJL_Figure_Delta}
\end{figure}

The results are shown in terms of gaps with specific
color-flavor symmetry in Fig. \ref{NJL_Figure_Delta}.
These results show that the $\Delta_{66}$ gaps are non-zero but
small. As well, the effect described above is seen more clearly:
significant disruption to the pairing of the heavy quark with the
light quarks in the dominant channel and a slightly smaller enhancement of the pairing
of the light quarks.  These results qualitatively agree with the
results of \cite{SW_Description} as far as the increase in
$\Delta_{\bar{3}\bar{3}}(ud)$ and decrease of in
$\Delta_{\bar{3}\bar{3}}(us)$.  The quantitative agreement appears at
first sight to be rather worse than above.  The differences between the results
of \cite{SW_Description} and \cite{Unlocking_ABR}, however, are 
mainly the result of different parameter choices\cite{S_Communication}.
Therefore the results presented here for the NJL model are
consistent with previous analyses. 

The mathematical reason for this phenomenon can be seen by examining
the gap equations for the individual $\Delta_{\bar{3}\bar{3}}$
gaps, but neglecting the $\Delta_{66}$ gaps and the
difference between the $\Delta_{\bar{3}\bar{3}}$ gaps on the
right hand side of the gap equation.  Expanding the integrands on the right
hand side of the gap equations one obtains:
\begin{eqnarray}
&\!\!\!\!\!\!\!\!\!\!\!\!\!\!\!\!\!& \Delta_{\bar{3}\bar{3}}(ud)=4 G\int \frac{dq} {2 \pi^2}\; q^2 \;{\cal R}(q)\left[
\frac{2 \meanDelta33}{9(l^2+\meanDelta33^2)^{1/2}}\right. +
\frac{ \meanDelta33}{9(l^2+4\meanDelta33^2)^{1/2}} 
\label{Delta_33(ud)_1GX}\\ &\!\!\!\!\!\!\!\!\!\!&
+
\left(\frac{ l (2l^2+5\meanDelta33^2)}{162\meanDelta33(l^2+\meanDelta33^2)^{3/2}} \right.
 -  \left.
\frac{ l (2l^2+14\meanDelta33^2)}{162\meanDelta33(l^2+4\meanDelta33^2)^{3/2}} \right)
\delta\mu \nonumber\\ &\!\!\!\!\!\!\!\!\!\!\!\!&
+ 
\left(\frac{ 8 l^6 + 28 l^4 \meanDelta33^2+29 l^2 \meanDelta33^4+18\meanDelta33^6}{486\meanDelta33^3(l^2+\meanDelta33^2)^{5/2}} \right.
 - \left.\left.
(\frac{ 8 l^6 + 88 l^4 \meanDelta33^2+314 l^2 \meanDelta33^4+396\meanDelta33^6}{486\meanDelta33^3(l^2+4\meanDelta33^2)^{5/2}} \right)
\delta\mu^2 \right] \nonumber \\ &\!\!\!\!\!\!\!\!\!\!&
  +  {\cal O}(\delta\mu^3) \nonumber
\end{eqnarray}
\begin{eqnarray}
&\!\!\!\!\!\!\!\!\!\!& \Delta_{\bar{3}\bar{3}}(us)=4 G\int \frac{dq} {2 \pi^2}\; q^2 \;{\cal R}(q)\left[
\frac{2 \meanDelta33}{9(l^2+\meanDelta33^2)^{1/2}}\right. + 
\frac{ \meanDelta33}{9(l^2+4\meanDelta33^2)^{1/2}} 
\label{Delta_33(us)_1GX}\\ &\!\!\!\!\!\!\!\!\!\!&
+
\left(-\frac{ l (l^2+7\meanDelta33^2)}{162\meanDelta33(l^2+\meanDelta33^2)^{3/2}} \right.
 +  
\frac{ l (l^2-2\meanDelta33^2)}{162\meanDelta33(l^2+4\meanDelta33^2)^{3/2}} 
-\left.
\frac{  l \meanDelta33}{12(l^2+\meanDelta33^2)^{3/2}}\right)
\delta\mu \nonumber\\ &\!\!\!\!\!\!\!\!\!\!&
+ 
\left(\frac{ 2 l^6 + l^4 \meanDelta33^2+ 8 l^2 \meanDelta33^4-9\meanDelta33^6}{486\meanDelta33^3(l^2+\meanDelta33^2)^{5/2}} \right.
 - 
\frac{ 2 l^6 + 16 l^4 \meanDelta33^2+14 l^2 \meanDelta33^4-36\meanDelta33^6}{486\meanDelta33^3(l^2+4\meanDelta33^2)^{5/2}}  \nonumber \\ &\!\!\!\!\!\!\!\!\!\!&
\left. \left. 
+\frac{ (2 l^2 -\meanDelta33^2)\meanDelta33}{48(l^2+\meanDelta33^2)^{5/2}}
\right)\delta\mu^2 \right]
  +  {\cal O}(\delta\mu^3) \nonumber
\end{eqnarray}

One expects these equations to be dominated by the region near the Fermi surface, $l=0\, (q=\mu)$.
Expanding about $l=0$ one obtains an approximation for the integrand in this region
of the integration:
\begin{equation}
\!\Delta_{\bar{3}\bar{3}}(ud): q^2 \;{\cal R}(q)\left[
\frac{5}{18} + \frac{13}{648}\frac{l \; \delta\mu}{\meanDelta33^2}+\frac{5}{432} \frac{\delta\mu^2}{\meanDelta33^2}\right] 
 + {\cal O}(\delta\mu^3/\meanDelta33^3) + {\cal O}(l/\meanDelta33)
\end{equation}
\begin{equation}
\!\Delta_{\bar{3}\bar{3}}(us): q^2 \;{\cal R}(q)\left[
\frac{5}{18} - \frac{83}{648}\frac{l \; \delta\mu}{\meanDelta33^2}- \frac{1}{27} \frac{\delta\mu^2}{\meanDelta33^2}\right] 
 + {\cal O}(\delta\mu^3/\meanDelta33^3) + {\cal O}(l/\meanDelta33).
\end{equation}
The $l>0$ part of this approximation is dominant because of the factor, $q^2=(l+\mu)^2$.
Therefore, if this region of the integration is dominant,
$\Delta_{\bar{3}\bar{3}}(ud)$ should increase linearly with $\delta\mu$
and a quadratic component for larger values of $\delta\mu$.  Similarly
$\Delta_{\bar{3}\bar{3}}(us)$ should decrease linearly with $\delta\mu$
and a quadratic component for larger values of $\delta\mu$.  The decrease
in $\Delta_{\bar{3}\bar{3}}(us)$ should be greater than the increase in
$\Delta_{\bar{3}\bar{3}}(ud)$. This is exactly the type of behavior exhibited 
by the solutions as shown in Fig. \ref{NJL_Figure_Delta}.

As was mentioned above, this behavior was already seen in a different
model in \cite{Unlocking_ABR}.  One reason for presenting these
results is as a check of the method to show that the model used
here gives results that are consistent with previous analyses.  The
second and more important reason is to contrast with the
results in the next section using perturbation theory 
and to explain the differences.

\section{Perturbative Single Gluon Exchange}

In the perturbative analysis, following the analysis of
\cite{1GExchange}, the gluon propagator:
\begin{eqnarray}
D_{\mu\nu}(q) = \frac{P^L_{\mu\nu}}{q^2-\Pi_L} + \frac{P^T_{\mu\nu}}{q^2-\Pi_T(q)}
- \xi \frac{q_\mu q_\nu}{q^2},  
\end{eqnarray}
is used in the weak coupling limit where:
\begin{eqnarray}
P^L_{\mu\nu}\approx \delta_{\mu 0}\delta_{\nu 0}, & \frac{q_i q_j}{q^4} \approx \hat{q}_i \hat{q}_j.
\end{eqnarray}
and using:
\begin{eqnarray}
\Pi_L= m_D^2 & \Pi_T=\frac{\pi}{4} i m_D^2 \;\frac{|q_0|}{|q|}
\label{DebyeLandau}
\end{eqnarray}
with:
\begin{equation}
m_D^2=\frac{N_f g^2 \mu^2}{2 \pi^2}
\end{equation}
\begin{equation}
g^2=4 \pi \left( \frac{12 \pi}{(11 N_c-2 N_f) \log{\left[\mu^2/\Lambda^2_{QCD}\right]}}\right).
\end{equation}

Evaluating the right hand side of the gap equations is more complicated in
this case.  The gluon propagator depends on the angle, $\theta$, between the
momenta of the scattered fermions and this integral must be done
numerically.  There is still one trivial angular integral that can be
done immediately.  As in the previous section, the analytic continuation
$q_0 \rightarrow -i\,q_4$ is done and the $q_4$ integration is done by
picking up the poles (\ref{pole1})-(\ref{pole5}) of the quasiparticle propagator
in the upper half plane.  The right hand side of the gap
equations is then reduced to integration over $q(\equiv |\vec q|)$ and one
angular integral, $\int d\cos\theta$, which can be done numerically.  This approach has the
advantage that the quasi-particles are automatically on the mass
shell.

The gluon propagator is not the same for each term since it must be
evaluated at the poles of the quasiparticle propagator.  The 
poles of the gluon propagator are ignored in this calculation.

As was done in the previous section and in \cite{1GExchange}
the contribution of the antiparticle gaps is ignored.  This is because
the antiparticle gap does not lead to a gap at the Fermi
surface and it's contribution in the gap equation is
suppressed by at least a factor of $1/\mu^2$.  As well,
in a recent paper \cite{RD_gauge} it was shown that the
anti-particle gap is further suppressed by an extra power of the
coupling in perturbation theory if the quasiparticles are
on the mass shell.

In order to make a numerical solution of this problem feasible we
assume that the dominant contribution to the gap equation comes from
the mean value of the $ud$ and $us$ gaps which are antisymmetric in
color and flavor:
\begin{equation}
\bar{\Delta}_{\bar{3}\bar{3}} = 
\frac{\Delta_{\bar{3}\bar{3}}(ud)+\Delta_{\bar{3}\bar{3}}(us)}{2}= \frac{1}{8}(a+h-3e) +\frac{1}{4}(c-f)
.
\end{equation}
The other gaps and the splitting between $\Delta_{\bar{3}\bar{3}}(ud)$
and $\Delta_{\bar{3}\bar{3}}(us)$ were evaluated after solving to
check that they were significantly smaller than
$\bar{\Delta}_{\bar{3}\bar{3}}$ and could safely be neglected.

The gap equation in this case is:
\begin{eqnarray}
&\!\!\!\!\!\!\!\!\!\!\!\!\!\!\!\!&\bar{\Delta}_{\bar{3}\bar{3}}(p)= \frac{g^2}{12\pi^2} \int_0^\infty dq \int_{-1}^1 dx \;\;
%\sum_{q_i} \lim_{q_4 \rightarrow q_i}
\stackrel{\mbox{Residue}}{q_4 \rightarrow q_i} \; F(q,\bar{\Delta}_{\bar{3}\bar{3}}(q),\delta\mu)
 \label{1gx_gap_equation} \\
&\!\!\!\!\!\!\!\!\!\!\!\!\!\!&
\left[ \frac{3/2-1/2 x}{p^2+q^2-2 p\, q x+
[
%\frac{\pi}{4} m_D^2\frac{|p_4-q_4|}{|\vec{p}-\vec{q}|}
\Pi_T-(p_4-q_4)^2]} 
+ \frac{1/2+1/2 x}{p^2+q^2-2 p\, q x+
[ \Pi_L-(p_4-q_4)^2]} \right] 
%\stackrel{\mbox{Residue}}{q_4 \rightarrow q_i} F(q,\bar{\Delta}_{\bar{3}\bar{3}}(q),\delta\mu)&
\nonumber
\end{eqnarray}
where $q_i$ are the poles of the quasiparticle propagator, (\ref{pole1})-(\ref{pole5})
and $x\equiv\cos{\theta}$.
\newpage
\begin{eqnarray}
&& \!\!\!\!\!\!\!\!\!\!
F(q,\bar{\Delta}_{\bar{3}\bar{3}}(q),\delta\mu)= 2\pi  \left[ 
\frac{\meanDelta33}{8\sqrt{l^2+\meanDelta33^2}} + \frac{\meanDelta33}{12\sqrt{(l-\delta\mu/2)^2+\meanDelta33^2}}
\right.  \label{totalexpression}\\
&+& \frac{\meanDelta33 \left( (2 l + \delta\mu)\delta\mu + 7 \meanDelta33^2 +D(q,\bar{\Delta}_{\bar{3}\bar{3}}(q),\delta\mu)\right) }{24\sqrt{2} D(q,\bar{\Delta}_{\bar{3}\bar{3}}(q),\delta\mu) \left( 2 l^2 - 2l \delta\mu + \delta\mu^2+5 \meanDelta33^2+D(q,\bar{\Delta}_{\bar{3}\bar{3}}(q),\delta\mu)\right)}
\nonumber \\ 
&-& \left. \frac{\meanDelta33 \left( (2 l + \delta\mu)\delta\mu + 7 \meanDelta33^2 -D(q,\bar{\Delta}_{\bar{3}\bar{3}}(q),\delta\mu)\right) }{24\sqrt{2} D(q,\bar{\Delta}_{\bar{3}\bar{3}}(q),\delta\mu) \left( 2 l^2 - 2l \delta\mu + \delta\mu^2+5 \meanDelta33^2-D(q,\bar{\Delta}_{\bar{3}\bar{3}}(q),\delta\mu)\right)} \right]
\nonumber
\end{eqnarray}

\begin{eqnarray}
D(q,\bar{\Delta}_{\bar{3}\bar{3}}(q),\delta\mu)= \sqrt{9 \meanDelta33^4 + 2\meanDelta33 ^2 \delta\mu(2 l + 3\delta\mu)+\delta\mu^2 ( 2l -\delta\mu)}
\end{eqnarray}

%Expression (\ref{totalexpression}) comes from the integrands of (\ref{NJL_gap_equation_e})-%(\ref{NJL_gap_equation_c}) in the combination:
%\begin{equation}
%\frac{1}{8}(a+h-3e) +\frac{1}{4}(c-f)
%\end{equation}

One can solve the gap equation (\ref{1gx_gap_equation})
iteratively using numerical integration with some care.  In particular
the collinear singularity at $|p|=|q|$ and $x=1$ in the gluon
propagator must be handled.  In addition there are other features of
the integrand near $x=1$ which are nonsingular but must be handled with a little care.

Perturbative analysis is certainly reliable at asymptotic
densities where the Fermi momentum is very high.  If color
superconducting quark matter is actually observed, however, it will be
at moderate densities of 4-5 times nuclear matter density in the core
of neutron stars.  At $\mu=500$ MeV where the results below are obtained,
the perturbative single gluon exchange interaction should be considered a model which is
distinct from and can be compared to the NJL model which is a low
energy model.

\begin{figure}[ht]
\epsfysize=3.5in
\epsfbox[ 40 49 401 291]{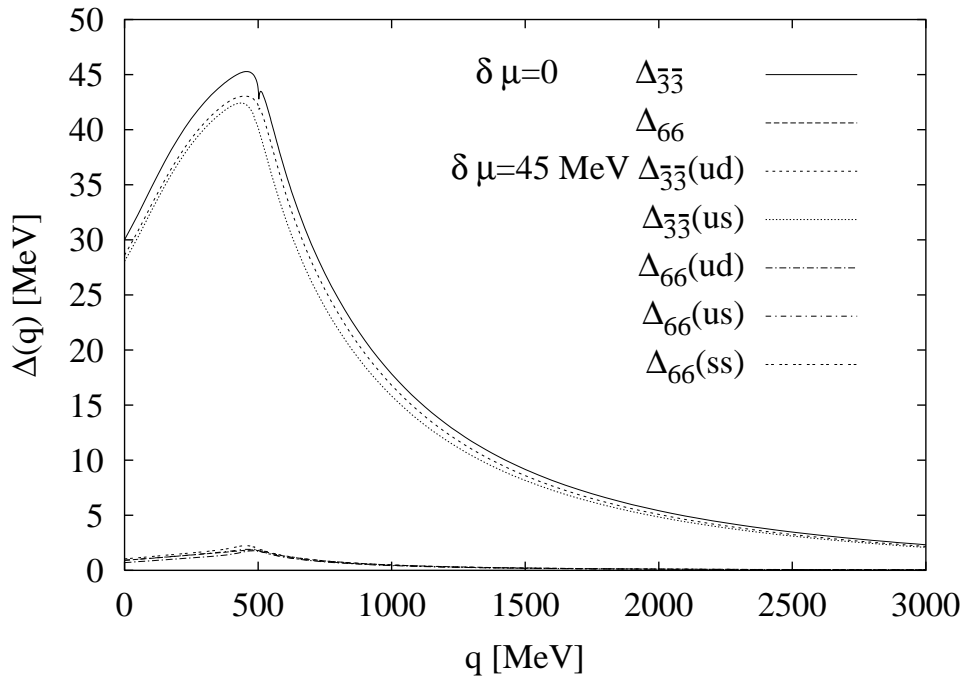}
   \caption{Numerical solutions for the gaps as a function of $p$ for $\delta\mu=0$ MeV
and $\delta\mu=45$ MeV.}
   \label{1GX_dmu0_dmu45_Figure}
\end{figure}

Using this procedure a solution of (\ref{1gx_gap_equation}) for
$\meanDelta33(p)$ was obtained for different values of
$\delta\mu$. These results were then used to calculate estimates for
the individual gaps.  The results are shown for $\delta\mu=0$ MeV and
$\delta\mu=45$ MeV in Figure \ref{1GX_dmu0_dmu45_Figure}.  The error
in the solution for $\meanDelta33$ is estimated to be less than $0.01
\%$ and the error in the values for the individual gaps are probably
less than $0.25 \%$.

Comparing the $\delta\mu=0$ results to previous analyses one sees that the
peak value of $\Delta_{\bar{3}\bar{3}}(p)$ is of the same order of
magnitude as predicted by the analysis of \cite{1GExchange} where
$\Delta_{\bar{3}\bar{3}}(p_0)$ is determined for the case when both
magnetic and electric gluon exchanges are taken into account. The form
of the $\delta\mu=0$ results is qualitatively similar to results found
in \cite{Kazu} for $\Delta_{\bar{3}\bar{3}}(p)$ in the two flavor
case at a similar density.

The agreement with \cite{Kazu} includes a small cusp feature seen most clearly in
the $\delta\mu=0$ results in Figure \ref{1GX_dmu0_dmu45_Figure}.  This feature arises because
the magnetic gluon propagator is non-analytic at $p_0=q_0$ due to the form
of $\Pi_T(p-q)\sim |p_0-q_0|$ in (\ref{DebyeLandau}).  This leads to
a non-analyticity of the gap function as a
function of $p$.  This feature arises in the results of
\cite{Kazu} as seen in their Fig. 6(a). 
It is a small effect, but the fact that it occurs in both analyses
supports the conclusion that the independent analyses are consistent.

The neglected $\Delta_{66}$ gaps are less than 5\% 
of the $\Delta_{\bar{3}\bar{3}}$ gaps, and $\Delta_{\bar{3}\bar{3}}(ud)$ and 
$\Delta_{\bar{3}\bar{3}}(us)$ are within 5\% 
of their mean value, $\meanDelta33$, at the maximum values of $\delta\mu$ investigated.  This 
indicates that the estimates obtained by solving for $\meanDelta33$ and then calculating
the individual gaps should be fairly reliable.

\begin{figure}[t]
\epsfysize=3in
\epsfbox[ 10 49 394 291]{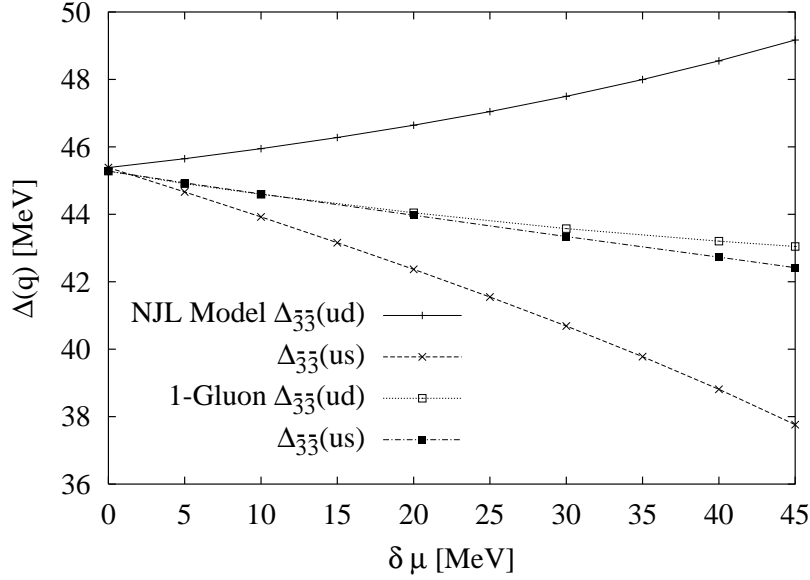}
   \caption{Comparison of the gaps as a function of $\delta\mu$ for the
NJL model and single gluon exchange.}
   \label{1GX_peak_Figure}
\end{figure}

\begin{figure}[ht]
\epsfysize=2.5in
\epsfbox[ 10 49 394 291]{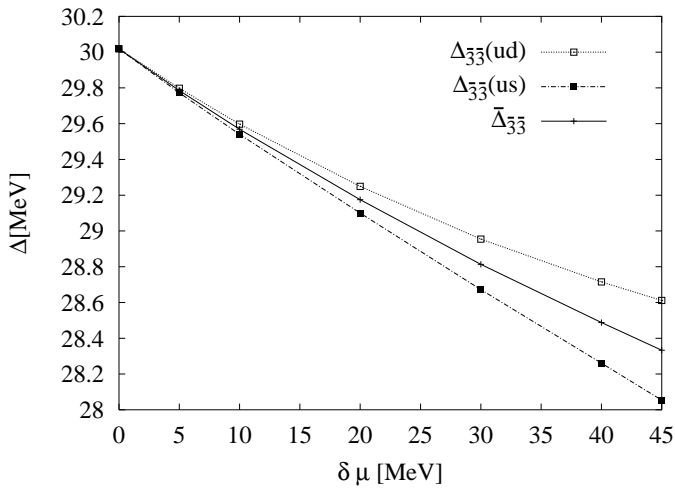}
   \caption{Gaps as a function of $\delta\mu$ in perturbation theory
at $p=0$.}
   \label{1GX_p0_Figure}
\end{figure}

\begin{figure}[ht]
\epsfysize=2.5in
\epsfbox[ 10 49 394 291]{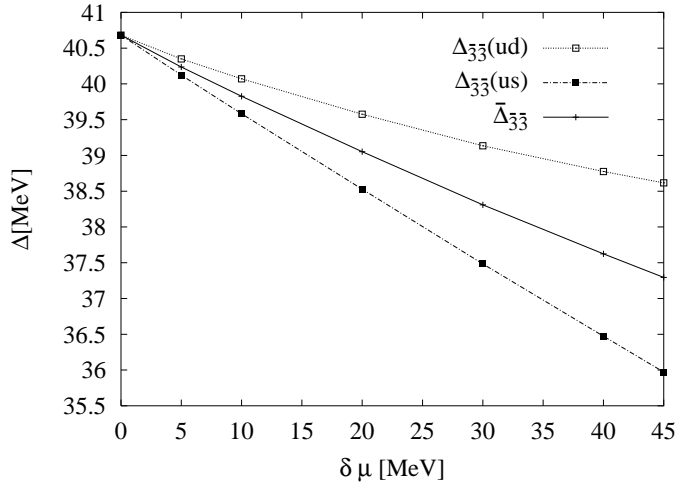}
   \caption{Gaps as a function of $\delta\mu$ in perturbation theory
at $p=550$.}
   \label{1GX_p550_Figure}
\end{figure}

\begin{figure}[ht]
\epsfysize=2.5in
\epsfbox[ 10 49 394 291]{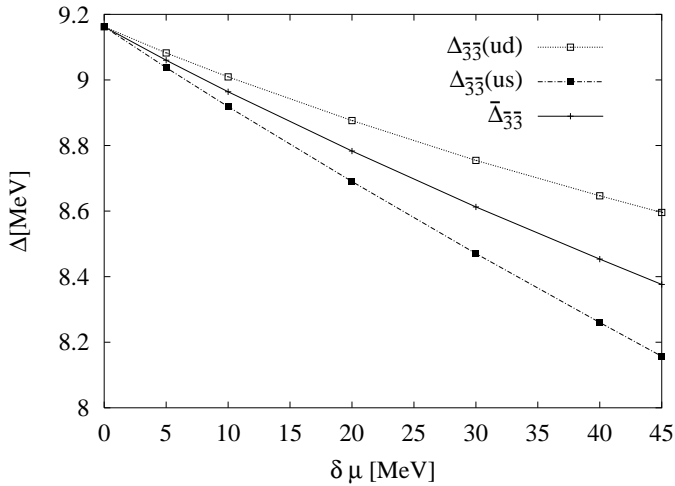}
   \caption{Gaps as a function of $\delta\mu$ in perturbation theory
at $p=1500$.}
   \label{1GX_p1500_Figure}
\end{figure}

The most important result of this analysis is that both
$\Delta_{\bar{3}\bar{3}}(ud)$ and $\Delta_{\bar{3}\bar{3}}(us)$
decrease with $\delta\mu$.  This is in contrast to the results
obtained in the NJL model where only the $\Delta_{\bar{3}\bar{3}}(us)$
gap decreases and the $\Delta_{\bar{3}\bar{3}}(ud)$ gap increases.
The peak value for these gaps is shown in Figure
\ref{1GX_peak_Figure} as a function of $\delta\mu$ with the NJL model
results from the previous section included for comparison.  
The results are also shown at 3 other fixed values of $p$ in Figures
\ref{1GX_p0_Figure}, \ref{1GX_p550_Figure} and \ref{1GX_p1500_Figure}.
These results show that both gaps consistently decrease with 
increasing $\delta\mu$ over the whole range of momenta and the 
decrease is basically linear in $\delta\mu$ with a small quadratic
component.


If one takes the objects in square brackets from the equations for $\Delta_{\bar{3}\bar{3}}(ud)$ and
$\Delta_{\bar{3}\bar{3}}(us)$ (\ref{Delta_33(ud)_1GX} and \ref{Delta_33(us)_1GX})
in the NJL model, substitutes into the perturbative equation 
and examines the region away from the peak at $l=0$
($l>>\meanDelta33$ but not asymptotic values of $l$), these terms are:

\begin{equation}
\Delta_{\bar{3}\bar{3}}(ud)(p): \left[
\frac{5}{18} - \frac{\meanDelta33 \, \delta\mu}{18 \: l^2} + {\cal O}(\delta\mu^2) \right] 
\label{1gx_gap_equation_ud}
\end{equation}
\begin{equation}
\Delta_{\bar{3}\bar{3}}(us)(p): \left[
\frac{5}{18} - \frac{5 \meanDelta33\, \delta\mu}{36 \: l^2} + {\cal O}(\delta\mu^2) \right] 
\label{1gx_gap_equation_us}
\end{equation}
Thus the effect of this region of the integration is to cause
both gaps to decrease linearly in $\delta\mu$ which is seen above.

The other effect that one observes from Figure \ref{1GX_peak_Figure} is that
the changes in the gaps in the NJL model are much more drastic than
in perturbation theory.  This is likely a result of the
fact that, with the cut-off in momentum space in the NJL model, the 
region of the integration near the Fermi momentum forms a much larger part
of the complete integral, and so the relative effect is larger.  In perturbation theory
where one integrates over a larger momentum region within which
different effects play a role, the relative change in the integral is much smaller.

Summarizing the results, at $\mu=500$ MeV the dependence of
$\Delta_{\bar{3}\bar{3}}(ud)$ and $\Delta_{\bar{3}\bar{3}}(ud)$ on
$\delta\mu$ in first order perturbation theory is much weaker than in
the NJL interaction model.  More importantly, the dependence of
$\Delta_{\bar{3}\bar{3}}(ud)$ on $\delta\mu$ actually changes sign.

The implications of this change in sign could be significant.  As
$\delta\mu$ increases, flavor is still locked to color except that $s$
flavor and $b$ color are distinguished from the other flavors and
colors.  Above $\delta\mu=2(\meanDelta33)_{\delta\mu=0}$ the two flavor
color superconducting phase is favored and the gap in this case is
larger than the $ud$ gap in the CFL phase.  If there is no intervening
phase between the three flavor superconducting phase and the two
flavor color superconducting phase, the fact that the $\Delta_{\bar{3}\bar{3}}(ud)$ gap decreases
suggests that the phase transition from the three to two flavor phases
is first order.  This analysis should be extended all the way to the
three to two flavor phase transition point in order to verify this.
As well, at this density, the inclusion of higher orders of
perturbation theory must be examined to determine how they affect the
results.  Finally this reasoning ignores the possibility of
intervening phases such as a three flavor equivalent 
of crystalline color superconductivity, which has been predicted in the
two flavor case\cite{Crystalline}, or CFL with meson condensation predicted in
\cite{Stress} in an effective Lagrangian model.  It would be very 
interesting to study both of these phenomenon in the present framework.

The results presented in this section were obtained using first order
perturbation theory at a moderate density where it should be
considered a model and may not be accurate.  One could speculate that
at much higher values of the chemical potential where perturbation
theory is definitely valid and the solution for $\meanDelta33$ is more
strongly peaked at $l=0$ \cite{Kazu}, the differences between the
perturbative results and the NJL model may not be as large as here.
However, these results indicate that the dependence on $\delta\mu$
should be weaker in perturbation theory than in the NJL model.

\section{Conclusion}

In this paper the general poles of the quasiparticle
propagator in the case where the $s$ quark has a different chemical potential
than the other two quarks has been presented.  These are a generalization of the poles
given in \cite{SW_Description}.  These poles are valid in a different
model than \cite{Unlocking_ABR} and the relationship between the two models
is not clear.

These poles were used in numerical solutions of the gap equations in
an NJL model to demonstrate that they produce results for the gaps which are
qualitatively consistent with \cite{Unlocking_ABR,SW_Description}.

The poles were also used in a numerical solution of the gap equation
for $\meanDelta33$ $(p)$ using perturbative single gluon exchange and
to obtain fairly reliable estimates of all the gaps.  The results
obtained for $\delta\mu=0$ are qualitatively and quantitatively
similar to those in \cite{Kazu}, which can be seen as support for both
results.  The results of \cite{Kazu} are obtained in the two
flavor case while the results presented here are obtained in the three flavor
case.

Finally. numerical results for the gaps were found as a function of
$\delta\mu$ out to $\delta\mu=45
MeV\approx(\meanDelta33)_{\delta\mu=0}$. The main conclusion is that
perturbation theory at $\mu=500$MeV predicts a decrease in
$\Delta_{\bar{3}\bar{3}}(ud)$ as a function of $\delta\mu$ where the
NJL model predicts an increase in $\Delta_{\bar{3}\bar{3}}(ud)$.
This observation suggests that the phase transition from
three to two flavor color superconductivity is first order.
The second conclusion is that the dependence of both
$\Delta_{\bar{3}\bar{3}}(ud)$ and $\Delta_{\bar{3}\bar{3}}(us)$ on
$\delta\mu$ is much weaker than in the NJL model for $\mu=500$MeV.
Therefore, there are both qualitative and quantitative differences
between the NJL model results and the leading perturbative
results at moderate density.  It is not clear which of these
results should be closest to the real world.

The weaker dependence of the gaps on $\delta\mu$ is likely to be true
even at much higher densities, but how much different perturbative
results might be from NJL results in this case remains to be
seen.  It is not clear if the increase of
$\Delta_{\bar{3}\bar{3}}(ud)$ with $\delta\mu$ will occur at higher
$\mu$ and this determination will be left for a future paper.

Analyzing the effect of shifting the chemical potential of the
strange quark by $\delta\mu$ is a first approximation to including
the effect of a strange quark mass.  This method has the benefit of separating
out the difficulties introduced by fully including a strange quark mass
in the gap equation.
%Knowledge of the new types of color-flavor structures 
%for condensates that arise when the $s$ quark is made distinct from the other 
%2 quarks at the level of the bare propagator.  
The knowledge gained in this research will be useful in an analysis where the 
$s$ quark is given a non-zero mass at the level of the bare quark propagator
which is the ultimate goal of this line of research\footnote{Recent 
papers \cite{Massive} have discussed progress in this direction.}.

As well, the dependence of the gaps on $\delta\mu$ presented
in this paper may be interesting when compared with
other papers that consider this effect.  In \cite{Diagrammatic}
it was shown in the two flavor case that up to $\delta\mu=\Delta_0/\sqrt{2}$,
$\delta\mu$ will have no affect on the gaps.  In \cite{Enforced} it is 
argued that in the three flavor case,
$\delta\mu$ will affect the values of the gaps and the
number densities of the quarks but the densities of the different
flavors of quarks will remain equal.  This situation would mean
that the superconducting ground state could remain electrically neutral
without the addition of electrons which would drastically change the
properties of this type of matter.  This conclusion could presumably 
be tested using the results of this paper.

This work can be extended in a number of ways.  The analysis should be
extended to $\delta\mu=2(\meanDelta33)_{\delta\mu=0}$ corresponding to the phase transition
to two flavors.  The analysis should also be done at higher densities
to determine how the results depend on $\mu$.  Determine of higher order corrections
to these results are necessary to determine the validity of the
perturbative approach at this density.  As well, the analysis of
this paper could be extended is to non-zero temperature
following \cite{RD_Weak}.

\section{Acknowledgments}

I would like to thank D. Rischke, R. Pisarski and T. Sch\"{a}fer for
many valuable discussions and K. Rajagopal for helpful comments.  This
research was funded by a Natural Science and Engineering Research
Council (NSERC) of Canada Post Doctoral Fellowship.  This work was
also supported in part by DOE grant DE-AC02-98CH10886.  Some of this
research was carried out during a visit to the Institut f\"{u}r
Theoretische Physik at the Johann Wolfgang Goethe Universit\"{a}t in
Frankfurt am Main and I would like to thank D. Rischke and the
Institut for their gracious hospitality.


\begin{thebibliography}{99}
%\bibitem{Barrois} B.C. Barrois, Nucl.Phys. {\bf B129} (1977) 390.
%\bibitem{BailinLove} D. Bailin and A. Love, Phys. Rep. {\bf 107} (1984) 325.
\bibitem{Barrois_BailinLove} B.C. Barrois, Nucl.Phys. {\bf B129} (1977) 390; \\
D. Bailin and A. Love, Phys. Rep. {\bf 107} (1984) 325.
\bibitem{ARW_RSSV} M. Alford, K. Rajagopal and F. Wilczek Phys.Lett. {\bf B422} (1998) 247; R. Rapp, T. Schaefer, E. Shuryak and M. Velkovsky, Phys.Rev.Lett. {\bf 81} (1998) 53.
\bibitem{Reviews} T. Sch\"{a}fer and E. Shuryak, nucl-th/0010049; \\
 K. Rajagopal and F. Wilczek, hep-ph/0011333.
\bibitem{CCSCCompact} M. Alford, J. Bowers and K. Rajagopal, J.Phys.{\bf G27} (2001) 541.
\bibitem{Carter_Blinking} G.W. Carter, {\it Color Superconductivity and Blinking Proto-Neutron Stars}, hep-ph/0111353;
G.W. Carter and S. Reddy, Phys.Rev. {\bf D62} (2000) 103002.
\bibitem{CFL} M. Alford, K. Rajagopal and F. Wilczek, Nucl.Phys. {\bf B537} (1999) 443.
\bibitem{CFL_accepted} T. Sch\"{a}fer and F. Wilczek, Phys.Rev.Lett. {\bf 82} (1999) 3956; \\
R. Rapp, T. Sch\"{a}fer, E.V. Shuryak and M. Velkovsky, Annals Phys. {\bf 280} (2000) 35;\\
 T. Sch\"{a}fer, Nucl.Phys. {\bf B575} (2000) 269; \\
I. Shovkovy and  L. Wijewardhana, Phys.Lett. {\bf B470} (1999) 189; \\ 
N. Evans, J. Hormuzdiar, S. Hsu and M. Schwetz, Nucl.Phys. {\bf B581} (2000) 391.
\bibitem{Crystalline} M. Alford, J. Bowers and K. Rajagopal, Phys.Rev. {\bf D63}
 (2001) 074016; K. Rajagopal, hep-ph/0109135.
\bibitem{Stress} P.F. Bedaque and T. Sch\"{a}fer, hep-ph/0105150; \\
P.F. Bedaque, nucl-th/0110049.
\bibitem{Unlocking_ABR} M. Alford, J. Berges and K. Rajagopal, Nucl.Phys.
{\bf B558}
(1999) 219.
\bibitem{SW_Description} T. Sch\"{a}fer and F. Wilczek, Phys.Rev. {\bf D60} 
(1999) 074014.
\bibitem{Asym_Bedaque} P. Bedaque, {\it Color Superconductivity in Asymmetric Matter}, hep-ph/9910247
\bibitem{Enforced} K. Rajagopal and F. Wilczek Phys.Rev.Lett. {\bf 86}
 (2001) 3492.
\bibitem{Diagrammatic}
J. Bowers, J. Kundu, K. Rajagopal and E. Shuster, Phys.Rev. {\bf D64} (2001) 014024.
\bibitem{Opening} A. Leibovich, K. Rajagopal and E. Shuster, Phys.Rev. 
{\bf D64} (2001) 094005.
\bibitem{1GExchange} T. Sch\"{a}fer and F. Wilczek, Phys.Rev {\bf D60} (1999) 114033.
\bibitem{RD_Weak} R. Pisarski and D. Rischke, Phys.Rev. {\bf D61} (2000) 074017.
\bibitem{Perturbation}  D.T. Son, Phys.Rev. {\bf D59} (1999) 094019; \\
R. Pisarski and D. Rischke, Phys.Rev. {\bf D61} (2000) 074017; \\
D. Hong, V. Miransky, I. Shovkovy and L. Wijewardhana, Phys.Rev. {\bf D61} (2000) 056001; Erratum-ibid. {\bf D62} (2000) 059903; \\
W. Brown, J. Liu and H. Ren, Phys.Rev. {\bf D61} (2000) 114012
\bibitem{Kazu} H. Abuki, T. Hatsuda and K. Itakura, {\it Structural Change of Cooper Pairs and Momentum-dependent Gap in Color Superconductivity}, hep-ph/0109013.
\bibitem{RD_superfluid} R. Pisarski and D. Rischke, Phys.Rev. {\bf D60} (1999) 
094013.
%\bibitem{MassTerms}  T. Sch\"{a}fer, {\it Mass Terms in Effective Theories of High Density Quark Matter}, hep-ph/0109052
\bibitem{S_Communication} T. Sch\"{a}fer, private communication.
\bibitem{RD_gauge} R. Pisarski and D. Rischke, {\it Gauge invaraince of the color superconducting gap
on the mass shell}, nucl-th/0111070.
\bibitem{Massive} M. Huang, P. Zhuang and W. Chao, hep-ph/0110046;
hep-ph/0112124.
\end{thebibliography}
\end{document}